\documentclass[12pt]{article}

\usepackage{fullpage}
\usepackage{amssymb, amsthm, amsmath}
\usepackage{doublespace}
\usepackage{bm}
\usepackage{graphicx}
\usepackage[authoryear]{natbib}
\usepackage{bm}
\usepackage{verbatim}
\usepackage{lineno}
\usepackage{times}
\usepackage{tabulary}
\usepackage{url}
\usepackage{gensymb}
\usepackage{subcaption}
\usepackage{booktabs}
\usepackage{authblk}

\newcommand{\beq}{ \begin{equation}}
\newcommand{\eeq}{ \end{equation}}
\newcommand{\beqn}{ \begin{eqnarray}}
\newcommand{\eeqn}{ \end{eqnarray}}

\title{Constrained Bayesian Nonparametric Regression
for Grain Boundary Energy Predictions}
\author[1]{Haoyu Wang}
\author[2]{Srikanth Patala}
\author[1]{Brian J. Reich}
\affil[1]{Department of Statistics, North Carolina State University}
\affil[2]{Department of Materials Science and Engineering, North Carolina State University}

\begin{document}
\maketitle

\begin{abstract}\begin{singlespace}
\noindent Grain boundary (GB) energy is a fundamental property that affects the form of grain boundary and plays an important role to unveil the behavior of polycrystalline materials. With a better understanding of grain boundary energy distribution (GBED), we can produce more durable and efficient materials that will further improve productivity and reduce loss. The lack of robust GB
structure-property relationships still remains one of the biggest
obstacles towards developing true bottom-up models for the behavior of
polycrystalline materials. Progress has been slow
  because of the inherent complexity associated with the structure of
  interfaces and the vast five-dimensional configurational space in
  which they reside. Estimating the GBED is challenging from a statistical perspective because there are not direct measurements on the grain boundary energy. We only have indirect information in the form of an unidentifiable homogeneous set  of  linear  equations. In this paper, we propose a new statistical model to determine the GBED from the microstructures of polycrystalline materials.  We apply spline-based regression with constraints to successfully recover the GB energy surface. Hamiltonian Monte Carlo and Gibbs sampling are used for computation and model fitting.  Compared with conventional methods, our method not only  gives more accurate predictions but also provides prediction uncertainties. \vspace{12pt} 
\end{singlespace}\end{abstract}
\newpage
\section{Introduction}\label{s:intro}

While the role of the structure of grain boundaries (GBs) in various
transport and failure mechanisms in polycrystalline materials has been
investigated for more than half a century \citep{forsyth1946grain,
  smoluchowski1952theory, haynes1955grain, hirth1972influence,
  hunderi1973influence, chadwick1976grain, gleiter1981interaction,
  gleiter1982structure, dimos1990superconducting,
  sutton1995interfaces, gottstein2009grain}, the lack of robust GB
structure-property relationships still remains one of the biggest
obstacles towards developing true bottom-up models for the behavior of
polycrystalline materials \citep{panchal2013key}.  This is
  because of the inherent complexity associated with the structure of
  interfaces and the vast five-dimensional configurational space in
  which they reside \citep{morawiec2003orientations,
    patala2012improved, patala2013symmetries}. Reliable
crystallography-structure-property relationships for interfaces are
particularly important for structural materials operating under
extreme environments, such as high temperatures, high strain rates and
dynamic loading conditions.

More recently, however, advances in both experimental and
computational techniques have facilitated large
databases of GB properties \citep{olmsted2009survey,
  olmsted2009survey2, holm2010comparing, rohrer2011grain,
  homer2014trends} in the five-parameter crystallographic
phase-space. The five macroscopic degrees of freedom (d.o.f) refer to
the misorientation (three parameters) and the boundary-plane
orientation (two parameters) aspects of the GB. With the advent of
modern high-throughput algorithms \citep{jain2011high,
  curtarolo2013high, jain2013commentary, saal2013materials} and
sophisticated experimental techniques \citep{seitahigh}, we have
reached a point where the development of new statistical tools is
critical for the analysis of the vast amounts of data being generated,
for developing novel scientific insights and for building
predictive models essential for the advancement of the field of GB
science and engineering.

One of the earliest high-throughput experimental techniques for the
measurement of GB properties is related to the relative energy
distributions of GBs in the five-parameter crystallographic
phase-space. Experimental measurements of GB energies rely on the
Herring equation \citep{herring1951some} that describes the equilibrium
condition of a triple-junction. For example, GB energies for copper
and aluminum were computed at high temperatures using the thermal
grooving measurements, where two free surfaces and a GB are in
equilibrium \citep{hasson1971interfacial,
  miura1994temperature}. Similarly, triple junction geometries are
determined to compute relative energies of experimentally observed
interfaces \citep{adams1999extracting, rollett2001grain}. The advent of
automated acquisition of large data sets of 3D EBSD data has
facilitated a sampling of triple junction
geometries to evaluate the relative energies of a large number of GBs
\citep{morawiec2000method}. Using this technique relative GB energies
have been computed for different structural metallic systems such as
nickel \citep{li2009relative}, aluminum \cite{barmak2005grain,
  barmak2006grain}, ferritic and austenitic steels
\citep{beladi2013relative, beladi2014five}; and ceramic materials
including magnesia (MgO) \citep{saylor2002distribution,
  saylor2000misorientation, saylor2003relative} and yittria
(Y$_2$O$_3$) \citep{dillon2009characterization,
  bojarski2013relationship, bojarski2012changes}. The statistical
distributions of different GB types (the GB character distribution)
and the relative energies are hosted online by Prof. Gregory
S. Rohrer's group at
 \url{http://mimp.materials.cmu.edu/~gr20/Grain_Boundary_Data_Archive/}. 
 %%% Haoyu's edit
\cite{morawiec2000method} presents a numerical method (referred to as block aggregation in this paper) for reconstructing the grain boundary energy distribution over the complete space of macroscopic boundary parameters. The method assumes that triple junctions are in local equilibrium, which is described by the Herring equation. The method discretizes the five-dimensional space and solves a homogeneous system of algebraic linear equations. 

In this paper, we propose a new nonparametric Bayesian model to reconstruct and predict grain boundary energy.  The method is based on generalized additive model (GAM) \citep{hastie2017generalized}, which is a generalized linear model with a linear predictor involving a sum of smooth functions of covariates. Each smooth function is defined by some basis functions, such as B-spline basis, polynomial basis and Gaussian basis.  GAMs have been proven to be extremely useful in analyzing data in complex domains \citep{walczak1996radial,
  scholkopf1997comparing, ramamoorthi2001efficient}. However, applying GAM to this problem is challenging because the physical properties of the GBs imply numerous constraints in the response surface that must be incorporated into the GAM model. In addition, these are not direct measurements on the grain boundary energy. We only have indirect information in the form of a homogeneous set  of  linear  equations.  We incorporate the constraints by implementing Hamiltonian Monte Carlo \citep{duane87} sampling and Gibbs sampling for posterior computation. The constraints enable the estimates of grain boundary energy identifiable.  Our constrained Bayesian nonparametric regression (CBNR) model outperforms the block aggregation (BA) method with respect to prediction accuracy. Our method also gives prediction intervals. This is the first time that GB energy uncertainties are quantified. 

The remainder of the paper is organized as follows. Section \ref{notation} introduces notation and the equations that define triple junctions. Section 3 presents our model and computational details. In Section 4, the method is compared with BA via a simulation study. Section 5 analyzes experimental data. Section 6 summarizes the paper and discusses future work.

\section{Analyzing the 3D EBSD Triple Junction Data}
\label{notation}
\begin{figure}[t]
		\centering
		\includegraphics[width=0.5\linewidth]{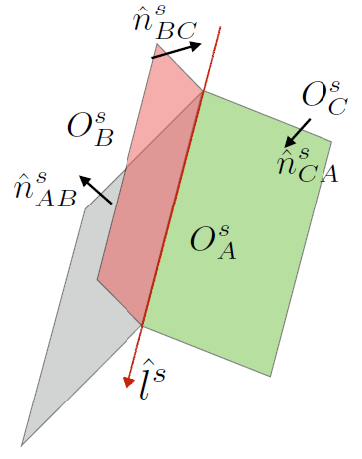}
        \caption{Illustrative example of one triple junction consisting of three grain boundaries: A (grey),B (red) and C (green). $\mathbf{O^s}$ is the grain orientation matrix, $\mathbf{\hat{n}^{s}}$ is the boundary-plane orientation vector and $\mathbf{\hat{l}^{s}}$ is the tangent vector for the triple junction line.}
\label{fig:fig1}
\end{figure}

The dataset has $n=19,094$ triple junctions. Figure \ref{fig:fig1} shows an example of one triple junction. For each triple junction, three grain boundaries are involved. For each grain boundary, the $3\times3$ grain orientation matrix $\mathbf{O}$, the $3\times1$ boundary-plane orientation vector $\mathbf{\hat{n}}$ and the $3\times1$ tangent vector for the triple junction line $\mathbf{l}$ are given in the dataset. The grain boundary misorientation matrix between grain boundary $\mathbf{b_i}$ and $\mathbf{b_j}$ is then defined as $\mathbf{M_{ij}} = \mathbf{O_i^{-1}O_j}$. The boundary-plane crystallography is thus defined by $\mathbf{M}$ and $\mathbf{\hat{n}}$. Further, the grain boundary misorientation matrix $\mathbf{M}$ can be transformed to a $3\times1$ vector $\mathbf{m}$, therefore, each grain boundary can be defined by five parameters: $\mathbf{b} = (\mathbf{m}, \mathbf{\hat{n}})$. We use the five-dimensional parameter $\mathbf{b}$ for the data analysis. Finally, the grain boundary $\gamma (\mathbf{b})$ is defined as
\[\gamma (\mathbf{b}) = \mathbf{\hat{n}(b)} \cdot \boldsymbol{\xi}(\mathbf{b}),\]
where $\boldsymbol{\xi}(\mathbf{b})$ is the $3\times1$ capillarity vector, which is unknown.

In addition, we consider two reference frames in this paper:
\begin{enumerate}
\item \emph{Crystal Reference Frame}: This refers to the Cartesian coordinate axes of the crystal (or grain) from which the vector, rotation matrices are described. We use a superscript $c$ to denote such quantities (e.g. $\mathbf{O^{c}}, \mathbf{\hat{n}^{c}}, \boldsymbol{\xi}^{c}$ etc.)

\item \emph{Lab/sample Reference Frame}: This refers to the fixed lab reference frame (usually fixed internally by the sample/detector geometry)
from which the vector, rotation matrices are described. We use a superscript $s$ to denote such quantities (e.g. $\mathbf{O^{s}}, \mathbf{\hat{n}^{s}}, \boldsymbol{\xi}^{s}$ etc.)
\end{enumerate}

\noindent The equation relating the capillarity vectors $\boldsymbol{\xi}$ and the tangent vector $\mathbf{t}$ of the junction

\begin{equation}
\left( \mathbf{\xi}^{s} \left( \mathbf{b}_1 \right) + \mathbf{\xi}^{s} \left( \mathbf{b}_2 \right) + \mathbf{\xi}^{s} \left( \mathbf{b}_3 \right) \right) \times \mathbf{l}^{s}=0
\label{Herring_samp}
\end{equation}

\noindent The energy of the grain boundary $\mathbf{b}_i$ is given by $\gamma (\mathbf{b}_i) = \mathbf{\xi}^{s} \left( \mathbf{b}_i \right) \cdot \mathbf{\hat{n}_i}^{s} = \mathbf{\xi}^{c} \left( \mathbf{b}_i \right) \cdot \mathbf{\hat{n}_i}^{c}$.

In order to apply the condition that the function defining the capillarity vectors is continuous, they have to be expressed in the crystal reference frame, i.e., we rewrite the capillarity vector in  (1) using the crystal reference frame. This is given by:
% We use the relationship $\vec{\xi}^c = O^s \vec{\xi}^s$ to re-write Equation \ref{Herring_samp} as:

\begin{equation}
\left( (\mathbf{O_1^s})^T\mathbf{\xi}^{c} \left( \mathbf{b}_1 \right) + (\mathbf{O_2^s})^T\mathbf{\xi}^{c} \left( \mathbf{b}_2 \right) + (\mathbf{O_3^s})^T\mathbf{\xi}^{c} \left( \mathbf{b}_3 \right) \right) \times \mathbf{l}^{s}=0
\end{equation}

\noindent We will not use superscript $c$ and $s$ representing reference frame for the rest of the paper. Our goal is to recover and predict grain boundary energies over the five-dimensional parameter space. Notice that the grain boundary energy is defined by the unknown capillarity vector $\boldsymbol{\xi}$ and known boundary-plane orientation vector $\mathbf{\hat{n}}$. Therefore, instead of modeling directly on grain boundary energy $\gamma(\cdot)$, we model on the capillarity vector $\boldsymbol{\xi}$. The challenge is that there are not direct measurements on either the grain boundary energy or the capillarity vector. We only have indirect information in the form of a homogeneous set of linear equations as in (1) and (2). In next section, we will present our model and computation details to solve this problem.

\section{Statistical model}\label{s:model}
\subsection{Model description and prior specification}
Using the notation defined in Section 2, for triple junction $i = 1,2,...,n$ we have
\begin{equation}
\left[ \mathbf{O_{i1}^T} \boldsymbol{\xi} \left( \mathbf{b}_{i1} \right) + \mathbf{O_{i2}^T} \boldsymbol{\xi} \left( \mathbf{b}_{i2} \right) + \mathbf{O_{i3}^T} \boldsymbol{\xi} \left( \mathbf{b}_{i3} \right) \right] \times \mathbf{l_i}=\mathbf{0},
\end{equation}
where $\mathbf{O_{ij}}$ is the $3\times 3$ grain orientation matrix, $\mathbf{b}_{ij}$ is the $5\times 1$ boundary-plane crystallography vector, and $\mathbf{l_i}$ is the $3\times 1$ tangent vector for triple junction $i$, $u \times v$ is the outer product of $u$ and $v$. The goal is to estimate the unknown capillarity vectors $\boldsymbol{\xi}(\mathbf{b}) = [\xi_1(\mathbf{b}),\xi_2(\mathbf{b}),\xi_3(\mathbf{b})]^T$. In matrix form, we can re-write (3) as
\begin{equation}
\mathbf{A_i} [\boldsymbol{\xi}( \mathbf{b}_{i1})^T,\boldsymbol{\xi}( \mathbf{b}_{i2})^T,\boldsymbol{\xi}( \mathbf{b}_{i3})^T]^T = 0,
\end{equation}
where $A_i$ is a $3\times 9$ matrix consisting of coefficients corresponding to the $i^{th}$ triple junction. Combining all the $n$ triple junctions, we have:
\begin{equation}
\mathbf{A} \boldsymbol{\xi} = 0,
\end{equation}
where $\mathbf{A}$ is a $3n \times 9n$ diagonal block matrix with diagonal blocks  $\mathbf{A_i}$ and  
\newline
$\boldsymbol{\xi} = [\boldsymbol{\xi}^T( \mathbf{b}_{11}),\boldsymbol{\xi}^T( \mathbf{b}_{12}),\boldsymbol{\xi}^T( \mathbf{b}_{13}),...,\boldsymbol{\xi}^T( \mathbf{b}_{n1}),\boldsymbol{\xi}^T( \mathbf{b}_{n2}),\boldsymbol{\xi}^T( \mathbf{b}_{n3})]^T$ is the $9n\times 1$ vector of unknown capillary.

%\subsection{Accounting for measurement error}
We assume the components $\mathbf{O_{ij}}$, $\mathbf{b}_{ij}$ and $\mathbf{l_i}$ are measured with error. Since it is impossible to estimate the errors associated with each component, we simply assume the random error model 
\[\mathbf{A}\boldsymbol{\xi} + \boldsymbol{\delta} = 0,  \]
where $\boldsymbol{\delta} \sim N(0, \sigma^2\mathbf{I})$.
This is equivalent to  the model
\[\mathbf{Y}|\boldsymbol{\xi} \sim N(\mathbf{A}\boldsymbol{\xi}, \sigma^2I),\]
where $\mathbf{Y}$ is a zero vector.

To model the underlying capillary process, suppose
\begin{equation}
	\boldsymbol{\xi}(\mathbf{b}) = f(\mathbf{b}) + \boldsymbol{\epsilon},
\end{equation}
where $\mathbf{f(b)} = (f_1(\mathbf{b}),f_2(\mathbf{b}),f_3(\mathbf{b}))^T$ and  $\boldsymbol{\epsilon} \sim N(\mathbf{0},\boldsymbol{\Sigma_{3\times 3}})$. The smooth function $\mathbf{f}$ can be approximated by any nonparametric basis functions, such as B-splines and Fourier functions, etc. For example, we  decompose $\mathbf{f}$ as the sum of main-effect functions: $f(\mathbf{b}) = \sum_{i=1}^{L=5} g_i(b_i)$, where $g_i(\mathbf{b}) = (g_{i1}(\mathbf{b}), g_{i2}(\mathbf{b}), g_{i3}(\mathbf{b}))$, is the $i^{th}$  additive main effect; here $L=5$ since $\mathbf{b}$ is five dimensional. Assuming the main-effect functions are sufficiently smooth, they can be approximated by B-spline basis expansions with B-spline basis functions $B_1(x),...,B_m(x)$. The main effect approximation is then $g_i(b_j) \approx \sum_{r=1}^{m} B_r(b_j)\beta_{ijr}$. Therefore, the regression of $\xi_i(\cdot)$, $i = 1,2,3$ can be modeled as 
\[	
	\xi_i(\mathbf{b}) = \sum_{j=1}^{5}\sum_{r=1}^{m} B_r(b_j)\beta_{ijr} + \epsilon_i \\
	= \mathbf{B(\mathbf{b})}\boldsymbol{\beta_i} + \epsilon_i.
	\]
The unknown coefficients are assigned non-informative normal priors, $\beta_{ijr} \sim N(0,\lambda)$ for large $\lambda$. 

The complete Bayesian hierarchical model is
\[\mathbf{Y}|\boldsymbol{\xi}, \boldsymbol{\beta},\sigma^2 \sim N(\mathbf{A}\boldsymbol{\xi}, \sigma^2I) \]
\[\boldsymbol{\xi} \sim N(\mathbf{B(\mathbf{b})}\boldsymbol{\beta},\boldsymbol{I}\otimes\boldsymbol{\Sigma}), \]
where $\otimes$ represents Kronecker product, $ \sigma^2 \sim InvGamma(a,b)$, $\Sigma \sim InvWishart(\Phi,df)$, and $\beta_{ijr} \sim N(0,\lambda), i=1,2,3, j=1,...,5, r=1,...,m$.  The hyperparameters $a$, $b$, $\Phi$, $df$ and $\lambda$ are set to give uninformative priors, as described in Section 3.3.

\subsection{Constraints}
In (3), the scale of $\boldsymbol{\xi}$ is not identified because if $\mathbf{A}\boldsymbol{\xi_0} = 0$, then $\mathbf{A}(c\boldsymbol{\xi_0})=0$ for any $c$. Also, depending on the rank of $\mathbf{A}$ the linear equation system may have infinitely many solutions and $\boldsymbol{\xi} = \mathbf{0}$ is always a solution. We impose the following constraints to ensure the capillary vector identified.
\begin{enumerate}
	\item Assume the $L_2$ norm of capillary vector is set to be greater or equal to a constant, that is, 
\begin{equation}
    ||\boldsymbol{\xi_{9n\times 1}}||_2 \geq D > 0,
\end{equation}
$D$ can be chosen arbitrarily, here we use $D = 9n$. 
\item Physical principles dictate that, for $\forall i \in \{1,2,...,n\}$, 
\begin{equation}
    \gamma(\mathbf{b_i}) = \hat{n}(\mathbf{b_i}) \cdot \boldsymbol{\xi}(\mathbf{b_i}) \ge 0.
\end{equation}
\end{enumerate}
Therefore, we have
$ ||\boldsymbol{\xi}||_2 \geq D$ and $ [\mathbf{C}\mathbf{B(\mathbf{b})}]\boldsymbol{\beta} \ge 0$, where $\mathbf{C}$ is the $3n \times 9n$ constraint matrix defined by the $\hat{n}(\mathbf{b_i})$ in the second constraint. The constraints clearly rule out the zero solution but do not fix the scale of $\boldsymbol{\xi}$. For this, we rescale the posterior of $\boldsymbol{\xi}$ as described in the next section.

\subsection{Posterior computation and model fitting}
We now describe the computational algorithm for our model. We sample the parameters using a combination of Gibbs sampling and Hamiltonian Monte Carlo sampling. The full conditional posterior distribution of $\boldsymbol{\beta}$ is truncated multivariate normal with mean $(\mathbf{B}^T\boldsymbol{\Omega^{-1}}\mathbf{B} + \boldsymbol{I}/\lambda^2)^{-1}\mathbf{B}^T\boldsymbol{\Omega^{-1}}\boldsymbol{\xi}$ and covariance $(\mathbf{B}^T\boldsymbol{\Omega^{-1}}\mathbf{B} + \boldsymbol{I}/\lambda^2)^{-1}$, constrained to $||\boldsymbol{\xi}||_2 \geq D, \mathbf{CB}\boldsymbol{\beta}\ge 0$. We sample $\beta$ using the Exact Hamiltonian Monte Carlo  \citep{pakman14}. Hamiltonian Monte Carlo (HMC) was first introduced by \cite{duane87}, where they united the MCMC and the molecular dynamics approach Hamiltonian dynamics \citep{alder59} to address lattice field theory simulations. Not long after, HMC began to be applied to statistical problems. Examples are \cite{neal1996, ishwaran1999, schmidt2009}. There have also been some tutorial and reviews on HMC such as \cite{neal1993, liu2008, neal2012}. \cite{pakman14} presented Exact HMC algorithm to sample from multivariate normal distribution in which the target space is constrained by linear and quadratic inequalities. 

We sample other parameters using Gibbs sampling. The complete MCMC sampling scheme follows the given process:
\begin{itemize}
	\item $\boldsymbol{\beta} | \boldsymbol{\xi},\lambda,\boldsymbol{\Omega} \sim N(\boldsymbol{\mu_1}, \boldsymbol{\Sigma_1})$, where $\boldsymbol{\mu_1} = (\mathbf{B}^T\boldsymbol{\Omega^{-1}}\mathbf{B} + \mathbf{I}/\lambda^2)^{-1}\mathbf{B}^T\boldsymbol{\Omega^{-1}}\boldsymbol{\xi}$ and $\Sigma_1 = (\mathbf{B}^T\boldsymbol{\Omega^{-1}}\mathbf{B} + \mathbf{I}/\lambda^2)^{-1}$
	\item $\boldsymbol{\xi} | \mathbf{Y},\sigma^2,\boldsymbol{\Omega},\boldsymbol{\beta} \sim N(\boldsymbol{\mu_2}, \boldsymbol{\Sigma_2})$, where $\boldsymbol{\mu_2} = (\mathbf{A^TA}/\sigma^2 + \boldsymbol{\Omega^{-1}})(\mathbf{A^TY}/\sigma^2 + \boldsymbol{\Omega^{-1}}\mathbf{B}\boldsymbol{\beta})$ and $\boldsymbol{\Sigma_2} = (\mathbf{A^TA}/\sigma^2 + \boldsymbol{\Omega^{-1}})^{-1}$
	\item $\sigma^2 | \mathbf{Y},\boldsymbol{\xi} \sim InvGamma(a+\frac{3n}{2}, b + \frac{(\mathbf{Y} - \mathbf{A}\boldsymbol{\xi})^T(\mathbf{Y} - \mathbf{A}\boldsymbol{\xi})}{2})$
	\item $\boldsymbol{\Sigma} |\boldsymbol{\xi},\boldsymbol{\beta} \sim InvWishart(\phi+\sum_{i=1}^{3n}\mathbf{a_i}, df+3n) $, where $\mathbf{a_i}_{3\times 3}$ is diagonal block matrix of $(\boldsymbol{\xi} - \mathbf{B}\boldsymbol{\beta})(\boldsymbol{\xi} - \mathbf{B}\boldsymbol{\beta})^T$
\end{itemize}
As described in section 3.2, the sclae of $\boldsymbol{\xi}$ is not identifiable. Therefore for each MCMC iteration, we compute $\tilde{\boldsymbol{\xi}}(\mathbf{b}) = \mathbf{B}(\mathbf{b})\boldsymbol{\beta}$ and rescale as $\boldsymbol{\xi}(\mathbf{b}) = \tilde{\boldsymbol{\xi}}(\mathbf{b}) / ||\tilde{\boldsymbol{\xi}}(\mathbf{b})||_2$. The capillary vector estimates are then the sample mean of the draws of $\boldsymbol{\xi}(\mathbf{b})$ and the pointwise $95\%$ prediction intervals are given by the sample quantiles of the MCMC draws of $\boldsymbol{\xi}(\mathbf{b})$.

\section{Simulation Study}\label{s:simulation}
We use one-dimensional and two-dimensional examples to illustrate our method in Sections 5.1 and 5.2, respectively. In Section 5.3, we conduct a five-dimensional simulation study to mimic the real data. 
For each simulation design we generate 100 datasets and compare our constrained Bayesian nonparametric regression (CBNR) model with block aggregation (BA). Methods are compared with respect to prediction mean squared errors (MSE) and 95\% credible set coverage (not available for BA). In the BA method, the parameter space is divided into discrete bins (so that each bin has 3 - 5 data points) and each bin is associated with one unknown capillarity vector $\boldsymbol{\xi}$. For every grain boundary in the experimental dataset, its capillarity vector is calculated by averaging all the capillarity vector of the bins that contain $\mathbf{b}$'s equivalences. To compare with the ground truth, we rescale the reconstructed responses so that the maximum is the same as the true maximum response. Prediction MSE is calculated using formula $MSE = \frac{1}{n}\sum_{i=1}^{n}  (\hat{\boldsymbol{\xi}}(\mathbf{b}) - \boldsymbol{\xi})^2$. Simulation results are shown in Table 1, both MSE and  95\% credible set coverage are the mean over the 100 simulated datasets.

\subsection{One-dimensional simulation}
First, we demonstrate our method using a one-dimensional example. Assume the true capillary function is 
		\[
		\xi(b) = 
		\begin{cases}
		\max [\sin(b),1/b ],& \text{if } b \in [0,2\pi]\\
		-5\sin(b/2),& \text{if } b \in [2\pi,4\pi]\\
		(b - 4\pi)^2 / 10 & \text{otherwise}.
		\end{cases}
		\]
The constraint here is $\sum_{i=1}^{n} \xi_i^2 \ge n$ and $\xi_i \ge 0$ for $i=1,2,...,n=100$. We assume we only know linear combinations of the corresponding responses are zeros. That is, we generate coefficient matrix $\mathbf{A}$, such that $\mathbf{A}\boldsymbol{\xi} = \boldsymbol{\delta}$, where $\boldsymbol{\xi} = (\xi_1,\xi_2,...,\xi_n)$, $\mathbf{b} \in [0,20]$, $\boldsymbol{\delta} \sim N(0,0.1)$ is random noise. To make $\mathbf{A}\boldsymbol{\xi} = \boldsymbol{\delta}$, we first generate random matrix $\mathbf{A}$ with standard normal elements. Denote the last column of $\mathbf{A}$ as $\mathbf{a}$ and the last element of $\boldsymbol{\xi}$ as $\phi$, we then replace the final column of $\mathbf{A}$ with $(\mathbf{a} - \mathbf{A}\boldsymbol{\xi} - \boldsymbol{\delta}) / \phi$. We denote this simulation as Case 1. Second, we generate 500 data points from one-dimensional mean-zero Gaussian process (GP) with Matern correlation function $cov(\xi_i, \xi_j) =\sigma^2(2^{\kappa-1}\Gamma(\kappa))^{-1} (||b_i - b_j||/\phi)^\kappa K_{\kappa}(||b_i - b_j||/\phi)$, where $\Gamma(\cdot)$ is the gamma function, $K_{\kappa}$ is the modified Bessel function of the third kind of order $\kappa$, $\phi$ is the range parameter and $\kappa$ is the smoothness parameter. We then generate coefficient matrix $\mathbf{A}$ as in Case 1. For the parameters used in generating Gaussian process data points, we set $\sigma^2 = 0.9$, $\phi= 0.1$ and $\kappa= 2$, and we do not include nugget.  We denote this simulation as Case 2. 

\begin{figure}[ht]
		\centering
		\includegraphics[width=1\linewidth]{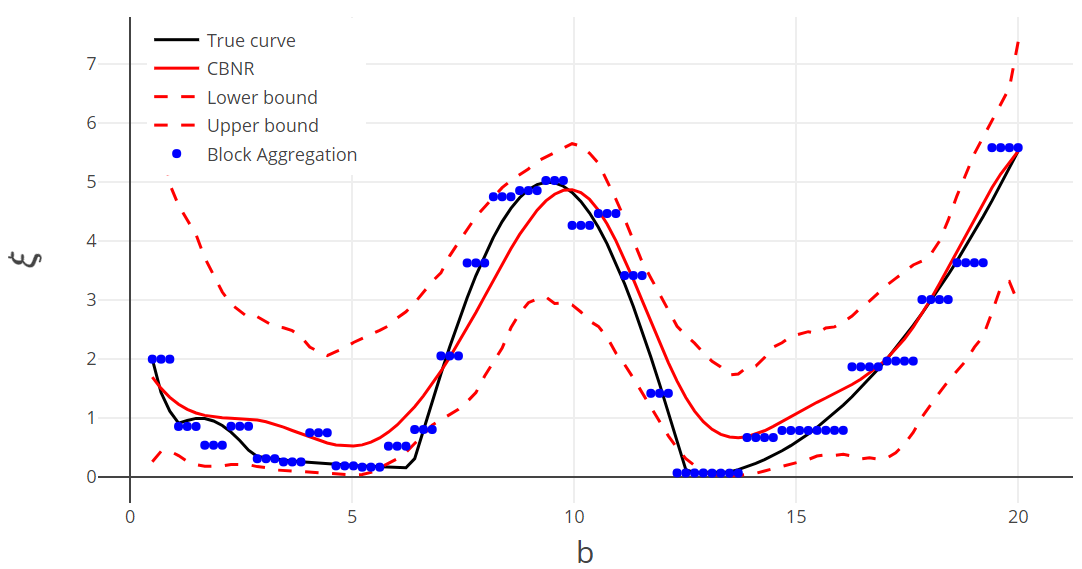}
        \caption{Analysis of one simulated dataset for Case 1. Plotted is the true curve $\boldsymbol{\xi}$ (black), the Constrained Bayesian Nonparametric Regression (CBNR) Method (red), $95\%$ credible set (dashed red) and the Block Aggregation method (blue).}
\label{fig:fig2}
\end{figure}

\begin{figure}[ht]
		\centering
		\includegraphics[width=1\linewidth]{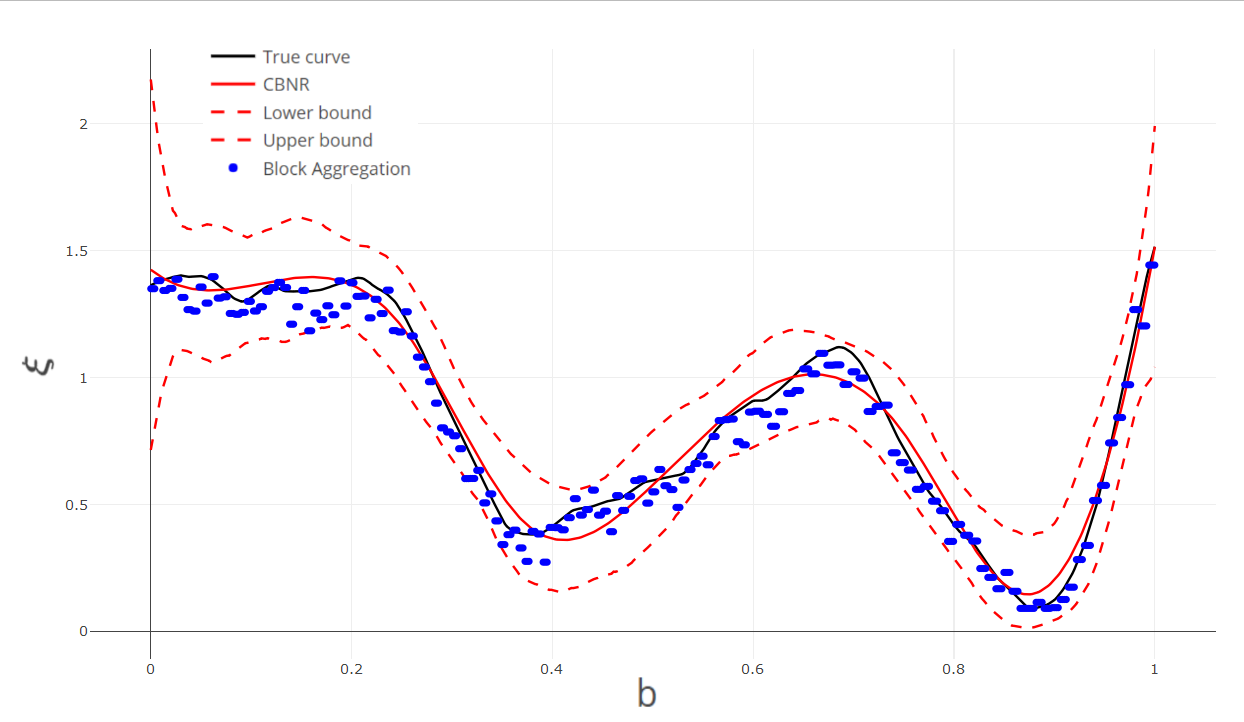}
         \caption{Analysis of one simulated dataset for Case 2. Plotted is the true curve $\boldsymbol{\xi}$ (black), the Constrained Bayesian Nonparametric Method (red), $95\%$ credible set (dashed red) and the Block Aggregation method (blue).}
\label{fig:fig3}
\end{figure}

Figures \ref{fig:fig2} and  \ref{fig:fig3} show the results of one simulated dataset for Case 1 and Case 2, respectively. Our CBNR model recovers the true curve smoothly because of the spline basis functions; the BA is not smooth, each bin shares the same estimated value. Note that CBNR can make predictions at new locations. However, since BA method uses numerical optimization to solve the linear equation system, it can only provide estimates at existing locations. In addition, according to Table 1, CBNR outperforms BA with MSE 3.6 times less for Case 1 and 6 times less for Case 2, which means CBNR gives more accurate predictions than BA. Furthermore, another advantage of CBNR over BA is that CBNR can provide prediction intervals. The 95\% CI coverage for Case 1 and 2 are 92.35\% and 92.51\%. 

\subsection{Two-dimensional simulation}
\begin{figure}[t]
\begin{subfigure}{.5\textwidth}
  \centering
  \includegraphics[width=0.9\linewidth]{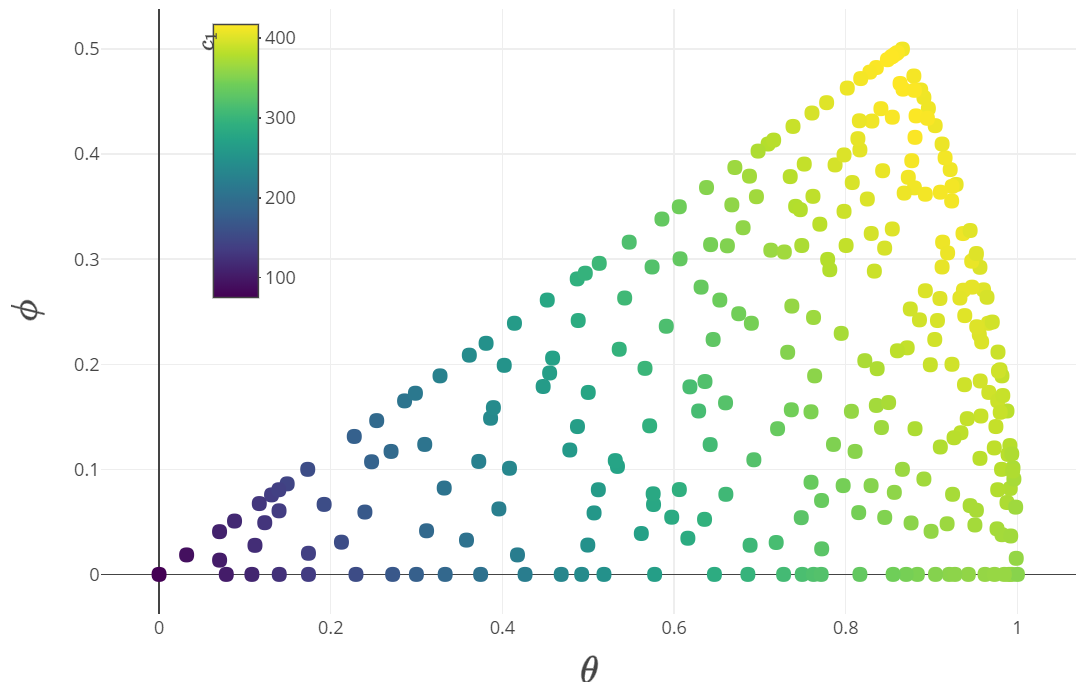}
  \caption{$C_1$ True Values}
  \label{fig:sfig3(a)}
\end{subfigure}%
\begin{subfigure}{.5\textwidth}
  \centering
  \includegraphics[width=0.9\linewidth]{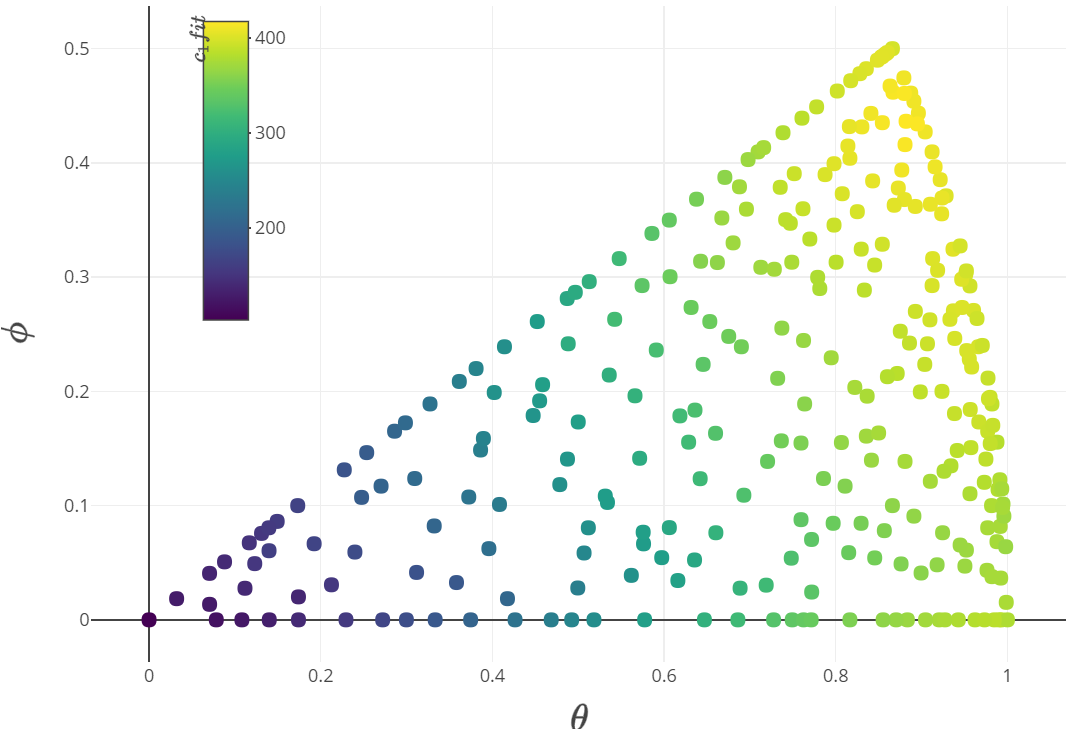}
  \caption{$C_1$ Fitted Values}
  \label{fig:sfig3(b)}
\end{subfigure}
\begin{subfigure}{.5\textwidth}
  \centering
  \includegraphics[width=.9\linewidth]{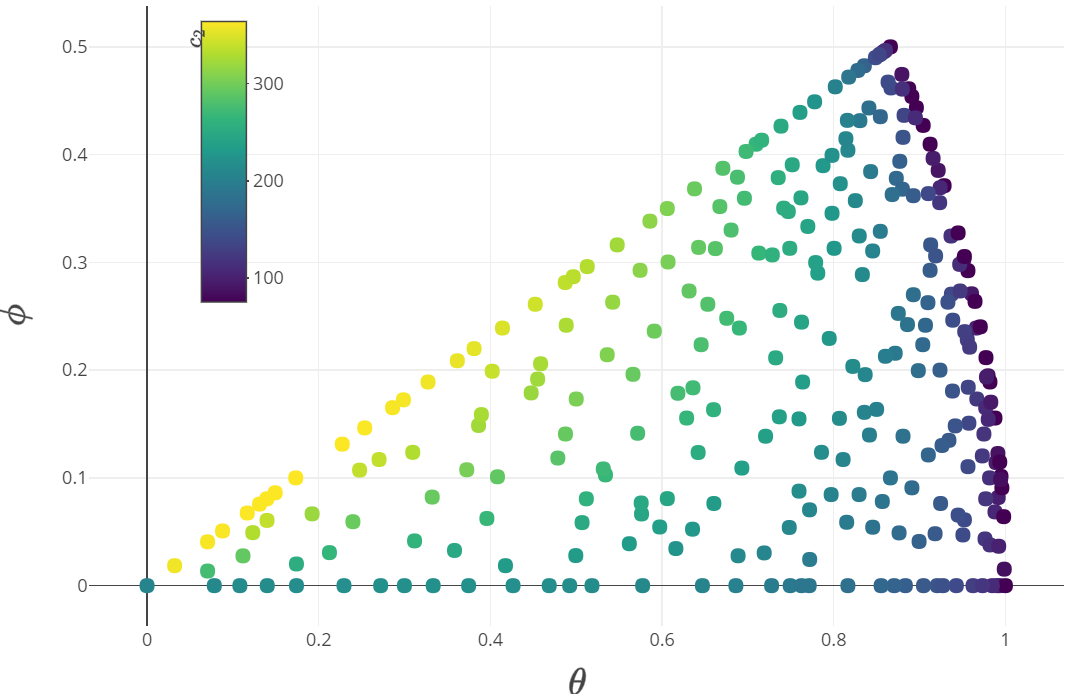}
  \caption{$C_2$ True Values}
  \label{fig:sfig3(c)}
\end{subfigure}%
\begin{subfigure}{.5\textwidth}
  \centering
  \includegraphics[width=.9\linewidth]{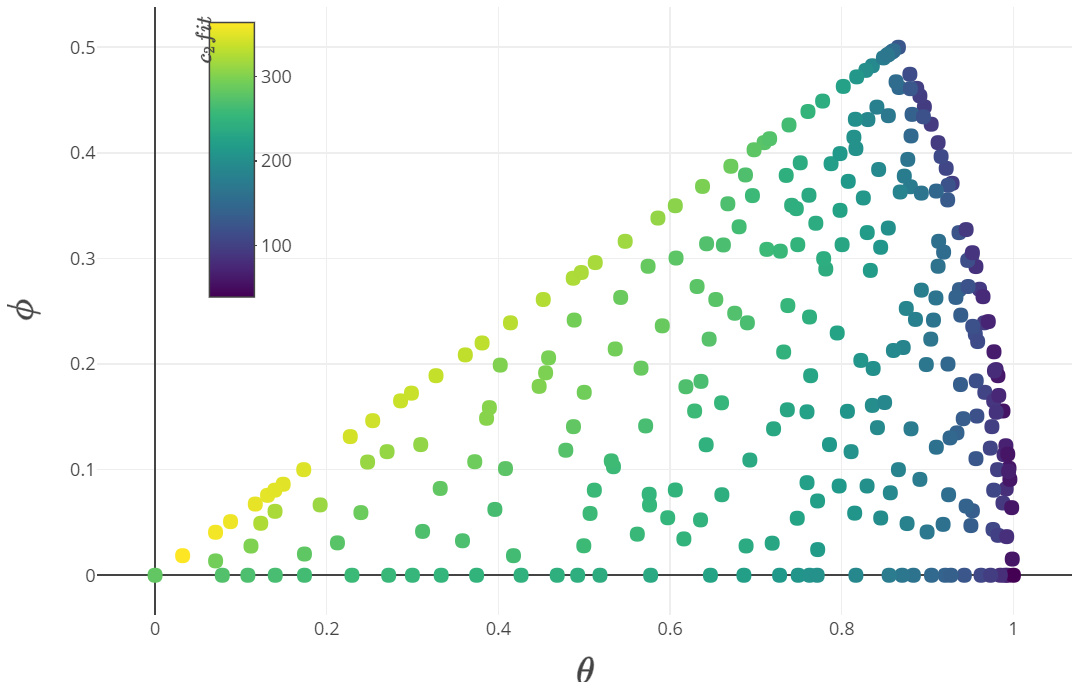}
  \caption{$C_2$ Fitted Values}
  \label{fig:sfig3(d)}
\end{subfigure}
\caption{Analysis of one simulated dataset for Case 3. Panels (a) and (c) show the true values and panels (b) and (d) show the recovered values by Constrained Bayesian Nonparametric Regression method.}
\label{fig:fig4}
\end{figure}

We then conduct simulations for two dimensional case. Assume the true curves are
		\[\begin{cases}
		c_1 = 75\cos\theta + 354\sin\theta\cos\phi + 206\sin\theta\sin\phi\\
        c_2 = 75\sin\theta + 354\cos\theta\sin\phi + 206\cos\theta\cos\phi
        \end{cases},\] 
and the constraints are $\sum_{i=1}^{294} (c_{1i}^2 + c_{2i}^2) \ge 294$, and $\theta_i c_{1i} + \phi_i c_{2i} \ge 0$, $i = 1,2,...,294$. Here, $\boldsymbol{\xi} = (c_1, c_2)$ and $\mathbf{b} = (\theta, \phi)$, $\theta \in [0,\pi/2]$, $\phi \in [0,\pi/6]$. As in the one-dimensional case,  we assume we only know linear combinations of the corresponding responses are zeros. Here,   we generate 100 different coefficient matrix $\mathbf{A}$, such that $\mathbf{Ac} = 0$, where $\mathbf{c} = (c_{1i},..c_{1n}, c_{2i},...,c_{2n}), n = 294$. We denote this simulation as Case 3. We project results on the sphere onto two-dimensional plane, Figure \ref{fig:fig4} shows the true values of $c_1$ and $c_2$, as well as fitted values on the 2-d plane, the CBNR model recovers the two-dimensional surfaces $c_1$ and $c_2$ precisely. Similar to one-dimensional case, our method has more accurate predictions than BA, with MSE being 86\% smaller than BA. In this case, the 95\% CI coverage is 99.65\%.

\subsection{Five-dimensional simulation}
Finally, we conduct a systematic simulation study based on real data. Since we suspect the capillarity vectors $\boldsymbol{\xi}(\cdot) = [\xi_1(\cdot),\xi_2(\cdot),\xi_3(\cdot)]^T$ in the application come from a smooth multivariate curve, to evaluate the performance of our method, we assume 
\[\boldsymbol{\xi}(\mathbf{b}) \sim GP(\boldsymbol{\mu},\boldsymbol{\Omega}),\]
where $\boldsymbol{\mu}$ is the mean of Gaussian process (GP), and $\Omega$ is the covariance matrix with exponential correlation function $cov(\boldsymbol{\xi_i}, \boldsymbol{\xi_j}) = \sigma^2\exp(-||\mathbf{b_i} - \mathbf{b_j}|| / \rho)$, where $\sigma^2$ is the variance, $\rho$ is the range parameter controlling spatial dependence and $||\cdot||$ is the Euclidean distance. Without loss of generality, we let $\boldsymbol{\mu} = 0$. Therefore, using the $5\times 1$ boundary-plane crystallography vector $\mathbf{b}$ in the real dataset, we generate capillarity vectors $\boldsymbol{\xi}$ from a GP with variance $\sigma^2 = 5$  and range parameter $\rho = 1$ and we do not add a nugget term. and coefficient matrix $\mathbf{A}$, such that $\mathbf{A}\boldsymbol{\xi} = 0$. We subsample 100 different sets of $\mathbf{b}$ in the real data. We denote this simulation as Case 4. As shown in Table 1, CBNR has prediction MSE 0.32, while MSE for BA is 1.59. In this case, the 95\% CI coverage drops down to 86.67\%. 

In conclusion, the simulation study shows three advantages of CBNR to BA. First, CBNR can make predictions at new data points, while BA can only recover at the existing points. Second, CBNR make much more accurate predictions than BA. Third, CBNR can provide prediction intervals, which BA can not offer.

\begin{table}[t]
\begin{center}
\begin{tabular}{ c|cc|c }
\hline
&\multicolumn{2}{c|}{CBNR} & BA\\
\hline
&MSE&$95\%$ CI Coverage&MSE\\
 Case 1 & 0.19 & $92.35\%$&0.87 \\ 
 Case 2 & 0.01  & $92.51\%$ & 0.08 \\ 
 Case 3 & 19.50 &$99.65\%$  & 136.42 \\ 
 Case 4 & 0.32 & $86.67\%$& 1.59\\ 
 \hline
\end{tabular}
\end{center}
\caption{Mean squared error and $95\%$ confidence interval coverage comparison between Constrained Bayesian Nonparametric Regression (CBNR) and Block Aggregation (BA) for simulated data. Case 1 and 2 refer to the one-dimensional simulations, Case 3 refers to the two-dimensional simulation and Case 4 refers to the five-dimensional simulation.}
\end{table}

\section{Real Data}\label{s:realdata}
In the dataset, there are $n = 19,094$ triple junctions. By imposing (7) in Section 3.2, all grain boundary energies can be determined up to a constant factor. Since we do not know the ground truth about grain boundary energy, metrics such as MSE can not be used to compare the two methods. However, according to Equation 5, the better method should have predictions with $\mathbf{A} \boldsymbol{\xi}$ closer to zero. Another factor we explore is the number of basis functions. We set number of basis functions to be $m=5,10,20,30$, and calculate $\frac{1}{3n}  \sum_{i=1}^{3n} (\mathbf{A_i} \hat{\boldsymbol{\xi}}(\mathbf{b}) - 0)^2$ for each $m$, where $\mathbf{A_i}$ is the $i^{th}$ row of coefficient matrix $\mathbf{A}$. We use five-fold cross validation to avoid overfitting.  Table 2 shows the relative prediction MSE of CBNR to BA. As the number of basis functions increases, the ratio decreases, meaning that more basis functions lead to better results. The ratio is always much smaller than one, meaning that our CBNR model outperforms BA with respect to prediction accuracy.

\begin{table}[t]
		\centering
		\label{tab:tab2}
		\begin{tabular}{ccccc}
			\hline
			No. of basis functions&5&10&20&30\\
			\hline
			Relative MSE&0.22&0.20&0.14&0.13\\
			\hline
		\end{tabular}	
		\caption{Relative prediction MSE of CBNR to BA}
	\end{table}

\begin{figure}[t]
		\centering
		\includegraphics[width=0.6\linewidth]{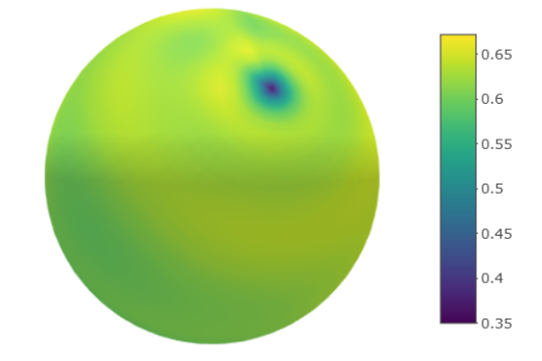}
         \caption{Stereographic projections of the grain boundary energy distribution for the $\Sigma3$ misorientation}
\label{fig:fig5}
\end{figure}

\begin{figure}[t]
		\centering
		\includegraphics[width=0.8\linewidth]{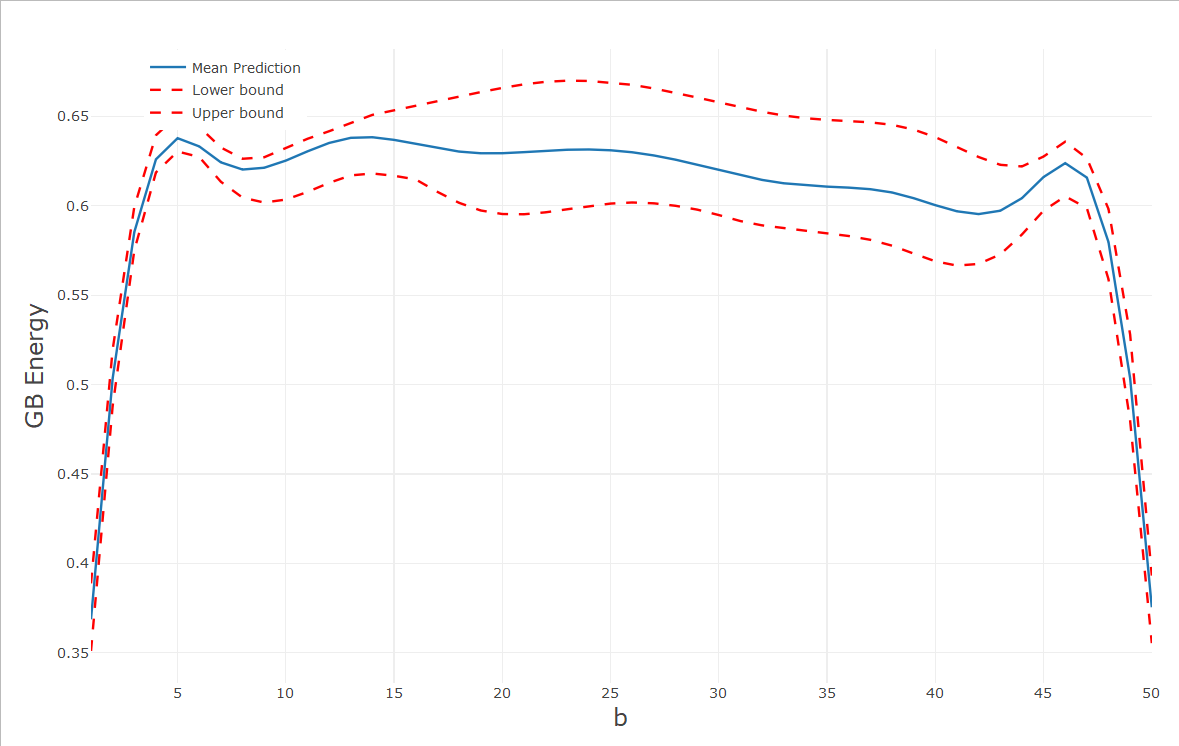}
         \caption{One dimensional prediction of the grain boundary energy distribution with $95\%$ prediction interval for the $\Sigma3$ misorientation fixing one boundary-plane parameters}
\label{fig:fig6}
\end{figure}

Visualizing the fitted surface of grain boundary energy in five-dimensional space is challenging.  One approach is to fix the three misorientation parameters and plot GB energy in the remaining two dimensions. One specific misorientation we are interested in is called $\Sigma3$, with misorientation parameters $(0.52,  0.96,  0.79)$. The two boundary-plane orientation parameters can be gridded on a 3D sphere, we predict GB energies and project them on the sphere. Figure \ref{fig:fig5} shows the results. The GB energy is strongly peaked at the position of the so-called coherent twin, which corresponds to boundary-plane parameters $(\arccos 1, \pi/4)$. This is due to the fact that the boundary-plane population is maximized at the coherent twin. Another advantage of CBNR against the conventional method BA is that we can provide prediction intervals. For instance, as shown in Figure \ref{fig:fig6}, we in further fix one of the two boundary-plane parameters and get one dimensional
predictions with $95\%$ prediction interval. the two rock bottom points at two sides correspond to the blue area the opposite side in Figure \ref{fig:fig5}. The prediction intervals are narrower than others because  boundary-plane population is maximized, in other words, we have more samples around these two areas than other areas.

Recall that we assume the components $\mathbf{O_{ij}}$, $\mathbf{b}_{ij}$ and $\mathbf{l_i}$ are measured with error and then assume a simplified random error model 
\[\mathbf{A}\boldsymbol{\xi} + \boldsymbol{\delta} = 0,  \]
now we conduct a simulation study to justify this model assumption. From the scientific fact we know that the measurement error associated with measuring the tangent vector $\mathbf{l_i}$ is larger than those with measuring $\mathbf{O_{ij}}$ and $\mathbf{b}_{ij}$. Therefore, we fix $\mathbf{O_{ij}}$, $\mathbf{b}_{ij}$ as in the real data and $\hat{\boldsymbol{\xi}}$ as estimated value. We then randomly generate error $\epsilon_{ij} \sim N(0, \gamma^2)$,$j = 1,2,3$ as the measurement error associated with $\mathbf{l_{ij}}$, $j=1,2,3$. In this way, we obtain the tangent vector with measurement error, and the element in the tangent vector is $\tilde{l}_{ij} = \frac{l_{ij} + \epsilon_{ij}}{\sqrt{\sum_{j=1}^{3} (l_{ij} + \epsilon_{ij})^2}}$. Denote the new coefficient matrix for the $i^{th}$ triple function as $\tilde{\mathbf{A}}_i$, we then calculate 
\[\boldsymbol{\delta}_i = (\tilde{\mathbf{A}}_i - \mathbf{A}_i) [\hat{\boldsymbol{\xi}}( \mathbf{b}_{i1})^T,\hat{\boldsymbol{\xi}}( \mathbf{b}_{i2})^T,\hat{\boldsymbol{\xi}}( \mathbf{b}_{i3})^T]^T.\]
We are interested in the empirical distributions of $\boldsymbol{\delta}_i$ with different amount of error variance $\gamma^2$. In addition, we would also like to see if the correlations between $\delta_{ij}$, $j=1,2,3$ are strong. We set $\gamma^2$ to be 0.01, 0.1, 0.05, 0.1, 0.5 and for each value of $\gamma^2$, we generate 100 replications and the results are shown in Table 3 and Figure 7.
From Table 3, even though the correlations increase as the variance $\gamma^2$ increases, the absolute values of these correlations are very small and close to zero. What's more, according to Figure 7, the histograms of $\delta_i$s under different values of $\gamma^2$ show that each of $\delta_{ij}$ is normal-like, with relatively small variances. The results from Table 3 and Figure 7 indicate that our assumption about the simplified random error makes sense.
\begin{table}[t]
		\centering
		\label{tab:tab3}
		\begin{tabular}{cccc}
			\hline
			$\gamma^2$&$r_{12}$&$r_{13}$&$r_{23}$\\
			\hline
		    0.01&0.0005(0.001)&-0.0002(0.001)&-0.001(0.001)\\
		    0.05&-0.001(0.001)&0.002(0.001)&0.0004(0.001)\\
		    0.1&.0.004(0.001)&0.005(0.002)&0.007(0.002)\\
		    0.5&-0.007(0.002)&0.02(0.002)&0.009(0.002)\\
			\hline
		\end{tabular}	
		\caption{Correlation between $\delta_{ij}$'s, where $r_{jk}$ is the correlation between $\delta_{ij}$ and $\delta_{ik}$.}
	\end{table}

\begin{figure}[t]
  \centering
  \begin{subfigure}[b]{0.3\linewidth}
    \includegraphics[width=\linewidth]{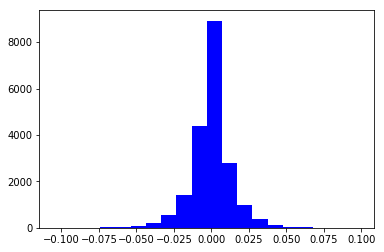}
     \caption{Histogram of $\delta_{i1}$, $\gamma^2 = 0.01$}
  \end{subfigure}
  \begin{subfigure}[b]{0.3\linewidth}
    \includegraphics[width=\linewidth]{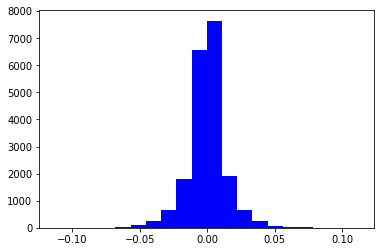}
    \caption{Histogram of $\delta_{i2}$, $\gamma^2 = 0.01$}
  \end{subfigure}
  \begin{subfigure}[b]{0.3\linewidth}
    \includegraphics[width=\linewidth]{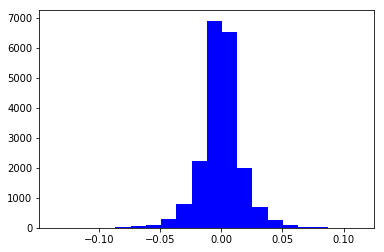}
    \caption{Histogram of $\delta_{i3}$, $\gamma^2 = 0.01$}
  \end{subfigure}
  \begin{subfigure}[b]{0.3\linewidth}
    \includegraphics[width=\linewidth]{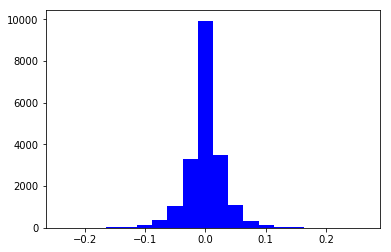}
     \caption{Histogram of $\delta_{i1}$, $\gamma^2 = 0.05$}
  \end{subfigure}
  \begin{subfigure}[b]{0.3\linewidth}
    \includegraphics[width=\linewidth]{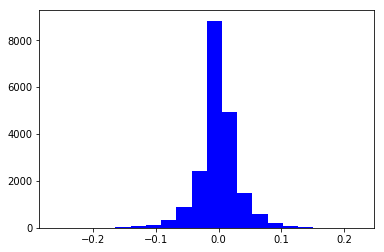}
    \caption{Histogram of $\delta_{i2}$, $\gamma^2 = 0.05$}
  \end{subfigure}
  \begin{subfigure}[b]{0.3\linewidth}
    \includegraphics[width=\linewidth]{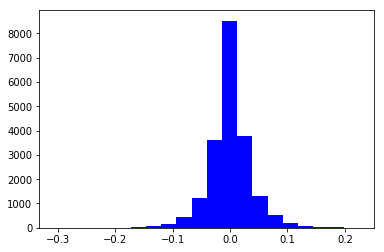}
    \caption{Histogram of $\delta_{i3}$, $\gamma^2 = 0.05$}
  \end{subfigure}
  \begin{subfigure}[b]{0.3\linewidth}
    \includegraphics[width=\linewidth]{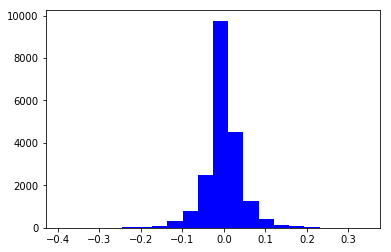}
     \caption{Histogram of $\delta_{i1}$, $\gamma^2 = 0.1$}
  \end{subfigure}
  \begin{subfigure}[b]{0.3\linewidth}
    \includegraphics[width=\linewidth]{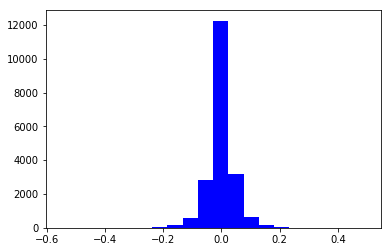}
    \caption{Histogram of $\delta_{i2}$, $\gamma^2 = 0.1$}
  \end{subfigure}
  \begin{subfigure}[b]{0.3\linewidth}
    \includegraphics[width=\linewidth]{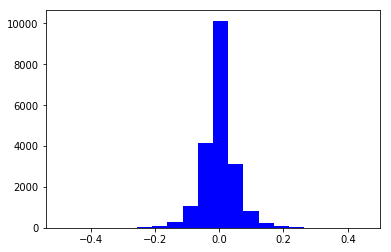}
    \caption{Histogram of $\delta_{i3}$, $\gamma^2 = 0.1$}
  \end{subfigure}
  \begin{subfigure}[b]{0.3\linewidth}
    \includegraphics[width=\linewidth]{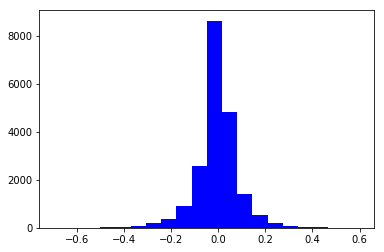}
     \caption{Histogram of $\delta_{i1}$, $\gamma^2 = 0.5$}
  \end{subfigure}
  \begin{subfigure}[b]{0.3\linewidth}
    \includegraphics[width=\linewidth]{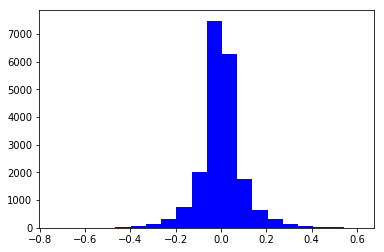}
    \caption{Histogram of $\delta_{i2}$, $\gamma^2 = 0.5$}
  \end{subfigure}
  \begin{subfigure}[b]{0.3\linewidth}
    \includegraphics[width=\linewidth]{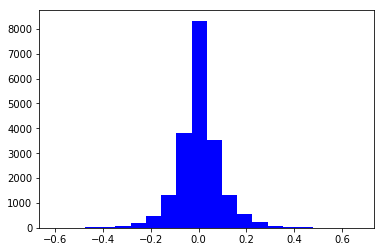}
    \caption{Histogram of $\delta_{i3}$, $\gamma^2 = 0.5$}
  \end{subfigure}
  \caption{Histograms of $\delta_i$ for different values of $\gamma^2$}
  \label{fig:fig7}
\end{figure}

\section{Discussion}\label{s:disc}
In this paper, we apply nonparametric Bayesian regression to grain boundary energy prediction. We formulate a new model that can recover and predict grain boundary energy relying on Herring's equation.  Our  method outperforms conventional numerical method with respect to prediction accuracy for both simulated and real data, and can provide prediction intervals. 

In geostatistics it is often desirable to take symmetry operations in the five-dimensional parameter space into account, as stated in Section 1. Therefore, a 
  capability of describing arbitrary functions in the five-parameter
  crystallographic phase-space of GBs will pave the way for applying
  statistical regression-based techniques and machine learning
  algorithms for the analysis of interface crystallography-structure
  and crystallography-property relationships. The numerical results are expected to be improved if we can extend the current spherical harmonics basis functions in the rotation space $SO(3)$ to the five-dimensional space.
  
%\section*{Acknowledgments}
%This work was partially supported by grants NSF-DMS-1513579, DOI-14-1-04-9, EPA-R835228, DE-AC02-05CH11231.

\begin{singlespace}
	\bibliographystyle{rss}
	\bibliography{GrainBoundary}
\end{singlespace}

\clearpage
\newpage
\section{Supplementary Information}

\subsection{Conventations and Notations}

% Where O^s_1, O^s_2, O^s_3 are from the triples.txt file! (They are given in passive notation)

In the \verb|triples.txt| data file:
\begin{enumerate}
    \item \textbf{Grain Orientations}:
    \begin{itemize}
        \item The orientations are specified in passive notation. Let's denote passive orientations it by $O^p$ and active orientations by $O^a$.
        \item The active and passive rotations are inversely related to each other, i.e. $O^a = \left( O^p \right)^{-1}$
        \item The misorientation will be defined as the active rotation of grain 2 with respect to grain 1. Therefore, $M_{12} = \left( O_1^a \right)^{-1} O_2^a$. If the orientations are provided in passive notation, $M_{12} = O_1^p \left(  O_2^p \right)^{-1} $.
    \end{itemize}
    \item \textbf{Boundary-plane normal vectors}:
    \begin{itemize}
        \item The components of the boundary-plane vector, $\hat{n}^s$, are expressed with respect to a fixed global reference frame.
        \item The vectors should be pointing in a clockwise direction with respect to the triple line.
        \item To convert the vector into the crystal reference frame, we refer to Figure \ref{fig:fig1}. 
        
        Suppose the orientations of grains $A$, $B$ and $C$, with respect to the global sample reference $(s)$, in active notation are $O_A^{s,a}$, $O_B^{s,a}$ and $O_C^{s,a}$ respectively. The boundary-plane vectors, with respect to the global sample reference $(s)$, are $\hat{n}_{AB}^{s}$, $\hat{n}_{BC}^{s}$, and $\hat{n}_{CA}^{s}$. The vector $\hat{n}_{ij}$ points from grain $i$ to grain $j$.
        
        Therefore, the vector $\hat{n}_{CA}^{c}$ is given by:
        
        $\hat{n} = \mathcal{B}_s n_s = \mathcal{B}_A n_A$, where $\mathcal{B}_A = \mathcal{B}_s O_A^{s,a}$

        $\hat{n} = \mathcal{B}_s n_s = \mathcal{B}_s O_A^{s,a} n_A$
        
        $n_s = O_A^{s,a} n_A$
        
        $n_A = \left[ O_A^{s,a} \right]^{-1} n_s$
        
        $n_A = O_A^{s,p} n_s$
    \end{itemize}
    \item \textbf{Boundary-plane normal vectors}:
    \begin{itemize}
        \item The components of the triple line vector are expressed with respect to a fixed global reference frame.
    \end{itemize}
    \item With respect to the basis functions, $\vec{xi} (\mathbf{b})$ has three components and each component is fit to a series of basis functions.
\end{enumerate}

Algorithm:

\begin{enumerate}
    \item Step 1: Compute all the $\mathbf{b} = (M; \hat{n})$ parameters.
    \item Step 2: The $\xi(\mathbf{b})$ is expressed in terms of the crystal reference frame. For the Herring equation, we need the components with respect to the sample reference frame.
    
    $\vec{\xi} =  \mathcal{B}_s \xi_s = \mathcal{B}_A \xi_A$
    
    $ \xi_s =  O_A^{s,a} \xi_A$
    
    $ \xi_s =  \left[ O_A^{s,p} \right]^{-1} \xi_A$
    
\end{enumerate}
\end{document}